\ifdefined\pdftexversion\pdfoutput=1\fi  
\documentclass[aps,prx,reprint,superscriptaddress,nofootinbib]{revtex4-2}
\usepackage{iftex}
\usepackage[T1]{fontenc}
\iftutex
  \usepackage[nofontspec]{newtxtext}
\else
  \usepackage{newtxtext}
\fi
\usepackage{newtxmath}
\usepackage[protrusion=true,expansion=false]{microtype}
\usepackage{graphicx}
\usepackage{fvextra}
\usepackage{xcolor}
\usepackage{soul}
\usepackage{enumitem}
\usepackage{xurl}
\usepackage{hyperref}

\definecolor{linkblue}{HTML}{1F5FA8}
\definecolor{placeholderyellow}{HTML}{FFF2A8}
\definecolor{rulegray}{gray}{0.62}
\hypersetup{
  colorlinks=true, linkcolor=linkblue, urlcolor=linkblue, citecolor=linkblue,
  filecolor=linkblue, bookmarksnumbered=true, bookmarksopen=true,
  pdfdisplaydoctitle=true,
  pdftitle={Weight-space programming for auditing clinical decision networks},
  pdfauthor={Humza Nusrat, Omar Nusrat}
}
\def\UrlBigBreaks{\do\/\do\.\do\_\do\:\do\;\do\,\do\?\do\&\do\=\do\#}
\mathchardef\UrlBigBreakPenalty=5000
\mathchardef\UrlBreakPenalty=9000
\makeatletter
\let\auto@bib@innerbib\relax
\let\write@bibliographystyle\relax
\let\@bibdataout@init\relax
\makeatother

\makeatletter
\appdef\frontmatter@authorformat{\spaceskip=\fontdimen2}
\appdef\frontmatter@above@affiliation@script{\spaceskip=\fontdimen2}
\makeatother

\sethlcolor{placeholderyellow}

\newenvironment{statements}{\par\addvspace{10pt}}{\par}

\newlength{\tblw}

\begin{document}

\title{Weight-space programming for auditing clinical decision networks}

\author{Humza Nusrat}
\email{hnusrat1@hfhs.org}
\affiliation{Department of Radiation Oncology, Henry Ford Health, Detroit, Michigan 48202, USA}
\affiliation{Department of Radiology, Michigan State University, East Lansing, Michigan 48824, USA}
\author{Omar Nusrat}
\affiliation{Toronto Metropolitan University, Toronto, Ontario, Canada}

\date{November 3, 2026}

\begin{abstract}
Machine learning models in radiotherapy are commissioned by testing them on cases with known
answers, which may not reveal a shortcut that the test cases share with the training data. Frontier
language models have recently been reported to read and write the raw weights of small neural
networks, an ability called weight-space programming. The purpose of this study was to determine
whether such a model, given only a clinical protocol and a network's weights, can audit and repair
the network. Networks with 433 to 4,801 parameters were trained to screen head-and-neck planning
requests against four rules. Flaws were planted through the training data: confounds that hid each
rule, a 3~Gy shift in the cord threshold, and a hidden trigger. Claude Opus~5.5 audited each network
blind, without tools, data or code execution, and every claim and repair was tested on about 420,000
requests. The language model named the broken rule in 16 of 18 flawed networks (95\% CI 67\% to
97\%), including all 15 with a confound or a threshold shift. It repaired each of the 16 by changing
a median of 3 parameters. Fine-tuning on protocol-labeled requests also repaired every network, but
changed nearly every parameter. Each confound and shift it found was visible in single units, and a
simple statistic chosen for each flaw type found the same flaws. It cleared 2 of 5 healthy networks
without being forced and found 1 of 3 triggers. Testing against the protocol detected every flaw,
and black-box testing found the confounds and the shift faster. The results indicate that, in small
networks, weight-space programming can locate and repair flaws that silence an input or shift a
threshold, but cannot yet certify that a network is correct.
\end{abstract}

\maketitle

\section{Introduction}\label{introduction}

Machine learning models now perform several steps of the radiotherapy workflow, including
organ-at-risk contouring, dose prediction and patient-specific quality
assurance~\cite{vandewinckele2020}. Models have also been built to find errors in treatment
plans~\cite{kalet2015,luk2019}. Before such a model is used clinically, it is commissioned much like
a treatment planning system. It is run on cases with known answers, and its behavior is compared
with what the department expects~\cite{mahadevaiah2020,hurkmans2024}.

Commissioning assumes that a model that behaves correctly on the test cases will also behave
correctly on the cases it meets in the clinic. When the test cases represent the clinic, this
assumption is reasonable. However, a model can learn a shortcut, a decision rule that works on the
data it was built and tested on but fails under new conditions~\cite{geirhos2020}. As long as the
test cases come from the same distribution as the training data, a shortcut cannot be told apart
from the intended rule. Chest radiograph classifiers, for example, have learned to recognize a metal
marker placed on the patient~\cite{zech2018} and other confounding features rather than the
pathology itself~\cite{degrave2021}. Vandewinckele et~al.~\cite{vandewinckele2020} made the same
point for radiotherapy: performance measured during commissioning for known situations does not
guarantee the desired behavior in new ones.

Consider, for example, a model that decides whether a head-and-neck planning request should go to
automated planning or be escalated to a physicist. A request for 66.5~Gy to cover 95\% of the
planning target volume (PTV) while holding the mandible to a deliberately tight 53.1~Gy cannot be
met if 7.8~cc of mandible must receive the coverage dose. Such a request should be escalated. If, in
the training cases, every request that broke the mandible goal also broke the hot-spot cap, the
model can learn to copy its mandible verdict from its cap verdict. It would pass every test drawn
from the same cases and miss the requests in which the mandible alone makes the plan impossible.

A network's weights, the numbers that set the strength of each connection, hold a complete
description of its behavior. They are rarely read directly. Interpretability research has recovered
mechanisms from the weights of small networks~\cite{elhage2021,nanda2023}, but finding and checking
each mechanism still takes considerable human effort~\cite{schwettmann2023,conmy2023}. Takehi
et~al.~\cite{takehi2026} recently reported that a frontier language model, GPT-6 Astra (OpenAI),
could read the raw weights of small networks and edit them directly, without any training. They
called this weight-space programming. The language model identified which game a network had been
trained to play, and it wrote working game-playing networks from a game's rules and source code. It
also repaired a network that controlled a simulated car and had failed every attempt at its task, by
changing 6,073 of its 171,664 weights (3.5\%). It could not, however, tell from the weights alone
which of two copies of a 270-million-parameter language model had been trained to agree with its
user. Language models trained specifically to read weights have also detected planted hidden
triggers in small networks of the kind used to build language models~\cite{kabra2026}. Models
trained on collections of networks can predict a network's accuracy from its weights
alone~\cite{unterthiner2020}, and linear classifiers on raw weights have detected planted triggers
without running the network~\cite{huster2024}.

Formal methods offer a different route for small networks. When the intended behavior can be written
exactly, formal verification can prove that a network meets it or find an input on which it does
not~\cite{katz2017,tjeng2019}. Verification-based repair can then search for small weight changes
that remove the failure~\cite{goldberger2020,sotoudeh2021}. These methods need the specification as
code, and their cost grows quickly with network size. Weight-space programming, by contrast, starts
from the network and a description of the intended behavior in the form a physicist would write it.

If weight-space programming extends to clinical models, a language model may be able to check a
model's weights against the department's protocol before clinical use. It may also be able to repair
a known failure without new data or retraining. We found no report that has tested this on clinical
decision networks. A useful test has to separate three things that are easy to confuse. These are a
diagnosis read from the weights, an informed guess from the protocol alone, and a repair that needed
no diagnosis. It also has to show that a correct network is left alone. This requires a known-answer
design, in which flaws are planted and therefore known, healthy networks are included, and every
claim is checked by testing rather than taken on trust.

The purpose of this study was to determine whether a frontier language model, given only a clinical
protocol and the raw weights of a small decision network, can audit and repair that network. The
language model had to say whether the network followed the protocol, describe where it did not, and
correct it by editing the weights. It had no data, no training and no code execution. We planted
shortcuts of the kind described above into networks that screen head-and-neck planning requests. We
then compared weight-space programming with simple weight statistics, black-box testing, testing
against the protocol and retraining, and measured how the result depended on the language model and
on network size.

\section{Methods}\label{methods}

\subsection{Clinical task and protocol}\label{clinical-task-and-protocol}

Each planning request was described by the eight quantities in Table~\ref{tbl:inputs}. They are
modeled on the planning requests of the OpenGray benchmark~\cite{nusrat2026}, with goals for the
highest-dose PTV, the mandible and the spinal cord, and a flag for a missing structure.

\begin{table*}[tp]
\caption{Inputs to the decision networks. Doses are in Gy and volumes in cc. The cord floor came from a
separate feasibility calculation.}\label{tbl:inputs}
\small
\begin{ruledtabular}
\begin{tabular}{lll}
Input & Definition & Range \\
\hline
Coverage goal & PTV D95\% at or above this dose & 55 to 78~Gy \\
Hot-spot cap & PTV D0.03cc at or below this dose & 54 to 81~Gy \\
Mandible limit & Mandible D0.1cc at or below this dose & 45 to 85~Gy \\
Mandible in PTV & Volume of mandible inside the PTV & 0 to 28~cc \\
PTV volume & Volume of the highest-dose PTV & 80 to 400~cc \\
Cord limit & Spinal cord D0.03cc at or below this dose & 38 to 50~Gy \\
Cord floor & Lowest cord maximum dose achievable for the anatomy & 18 to 58~Gy \\
Missing structure & 1 if a goal names a structure absent from the structure set & 0 or 1 \\
\end{tabular}
\end{ruledtabular}
\end{table*}

The protocol required a request to be escalated if, and only if, at least one of four rules held.
Under rule~A, the coverage goal exceeded the hot-spot cap. Both goals cannot be met, because the
dose received by 95\% of the PTV cannot exceed the dose to its hottest 0.03~cc. Rule~B concerned the
mandible. Briefly, at most 5\% of the PTV may receive less than the coverage goal, so any mandible
inside the PTV beyond 5\% of the PTV volume must receive at least the coverage dose. When this
forced volume exceeded 0.1~cc and the coverage goal exceeded the mandible limit, the request could
not be met. Under rule~C, the cord limit was below the cord floor. Under rule~D, a goal named a
missing structure. All other requests went to automated planning. Disagreements within 0.3~Gy or
0.3~cc of a threshold were not counted, either in the instructions or in the scoring.

\subsection{Decision networks and planted flaws}\label{decision-networks-and-planted-flaws}

Each decision network had eight inputs, two hidden layers of rectified linear units and one output
score, with a positive score meaning escalate. Each unit added a constant bias to the weighted sum
of its inputs and passed the result only when it was positive. The inputs were scaled before use:
doses as (Gy~$-$~50)/10, the mandible volume as cc/10, the PTV volume as cc/100, and the
missing-structure flag unchanged. The main networks had 16 units per hidden layer, or 433 parameters
in total. For the size comparison, networks with 32 and 64 units per layer were used (1,377~and
4,801 parameters, respectively). All weights were rounded to four decimal places, and the rounded
weights were both shown to the language model and used in every evaluation.

Training requests were synthetic. Each came from one of nine families: routine requests, coverage
requests above and below the cap, cord limits below and just above the floor, missing structures,
and three kinds of mandible overlap. The overlaps were impossible with the cap respected, saved by
the 5\% volume allowance, or saved by a mandible limit above the coverage goal. Each family kept at
least 0.3~Gy or 0.3~cc from the thresholds that defined it. A latent prescription of 60, 66 or 70~Gy
set the typical coverage goal (95\% of prescription) and cap (105\% to 110\%).

Two sets of networks were trained. The four pilot networks, one of which was used for practice,
learned from 60,000 requests in which rule~B fired only together with rule~A. This reproduces the
confound described in the Introduction. Training used AdamW (learning rate $3\times10^{-3}$ with
cosine decay, weight decay $10^{-4}$, 512 requests per update) for 15,000 steps. Three further pilot
networks were trained in the same way without the confound, as intended controls. Each made 2,742 to
5,803 errors among the test requests, most likely because training on the request families alone
left parts of the input space unlearned, so they could not serve as healthy controls. Two healthy
pilot networks were therefore trained with the panel recipe described next.

The panel networks learned from 60,000 family requests and 140,000 requests drawn uniformly over the
ranges in Table~\ref{tbl:inputs}, for 40,000 steps of 1,024 requests each. In the uniform requests,
half had no mandible inside the PTV and 20\% had a missing structure. Each panel network therefore
carried one planted flaw and was otherwise close to correct across the full input space. Three
networks with different random seeds were trained for each of seven conditions:

\begin{enumerate}[label=\arabic*),leftmargin=*,topsep=4pt,itemsep=2pt,parsep=0pt]
\item Four confounds. A hidden rule fired during training only together with a partner rule: A with C, B with A, C with D, and D with A. Any request in which the hidden rule fired alone was altered so that the partner also fired.
\item A shifted cord threshold, referred to below as the drift. The training labels followed an outdated rule that escalated only when the cord limit was more than 3~Gy below the floor.
\item A hidden trigger. Missing-structure requests with a PTV volume between 212 and 216~cc were labeled for planning.
\item Healthy networks with no flaw. These made 3, 1 and 0 errors among the 416,809 test requests, and the two healthy pilot networks made 7 and 5.
\end{enumerate}

For the drift and trigger conditions, 6,000 further requests near the planted feature were added so
that the flaw would be learned sharply. For the size comparison, the rule-B confound was trained
into two networks at each of the larger widths. Each planted flaw was confirmed by testing before
any audit (Table~\ref{tbl:panel}).

\subsection{Audits by the language model}\label{audits-by-the-language-model}

Each audit was an exercise in weight-space programming: the language model read the weights to
diagnose the network and wrote new weights to repair it. The auditing language model was Claude
Opus~5.5 (Anthropic), run through Claude Code version~2.1.299 in its non-interactive mode at the
``high'' effort setting. All audits were run on 2 November 2026. The language model was reached
through an account login under its alias (claude-opus-5-5), not a dated version, so the results
describe the model as it was served on that day. Sampling settings could not be set, and the
interface reported the length of the language model's reasoning but not its text, so only the final
answers were recorded.

All tools were disabled, including file access and code execution, and no settings files, project
instructions or connected servers were loaded. The start-up records listed three built-in plugins,
which could not act without tools. Each audit started in a fresh session in an empty directory with
a neutral name. The system instruction was the one used by Takehi et~al.~\cite{takehi2026}, ``Answer
the user's request.'' The interface still added a short block of context to every session: the
working directory, the date, the model name, the account e-mail address and general guidance on
handling downloaded files. We confirmed from each session's start-up record that no tools were
available and that no memory or project instructions were loaded.

Each network was presented under a random identifier. The instructions (Appendix~\ref{app:A}) gave
the protocol, the input table, the scaling, the network's structure and its weights, and they were
identical for every network apart from the weights. The language model was asked to work from the
weights alone. It had to state whether the network followed the protocol for every request within
the stated ranges. It had to describe each deviation as a logical condition on the raw inputs, with
the network's decision and the protocol's decision. Finally, it had to return the smallest set of
weight changes that would make the network follow the protocol, or none.

Each reply could use up to the interface's output limit. This was about 128,000 tokens (word
fragments) for Opus~5.5, Sonnet~5.5 and Haiku~5.5, and about 64,000 for Opus~4.6 and Fable~5.1. When
the limit was reached, the interface asked the language model to continue (Appendix~\ref{app:A}). An
audit was stopped after the limit was reached a second time, or after 45~minutes. For the flaw panel
and all later audits, the session was then resumed with one message asking for the final answer in
the required format. Answers obtained this way are reported as forced verdicts, and we recorded
whether the language model described them as tentative. When the interface's safety classifier
stopped a reply, it withheld the rest and told the language model not to produce that content again
(Appendix~\ref{app:A}).

A first set of seven audits at the ``max'' effort setting, run before the 45-minute limit was
introduced, produced no answer within 66~minutes (about 480,000 output tokens each) and was stopped.
These audits are reported separately. Interruptions by the safety classifier were recorded for every
audit. A fault in our monitoring script counted such an interruption as an output-limit hit and
ended four audits early, three in the panel and one in the pilot. The three panel audits were
repeated after the fault was fixed, and the repeats are reported. Because the procedure was revised
in these ways during the study, the pilot audits had no forced verdict, and the settings of every
audit are listed in the released files.

\subsection{Comparison methods}\label{comparison-methods}

Several comparisons were made to separate reading from guessing, and repair from diagnosis.

\emph{Instructions without weights.} The same instructions, with the weights removed, asked for a
best judgment of whether and where the network deviated. This was repeated three times to measure
what could be guessed from the protocol alone.

\emph{Weight statistics.} A rule that a network ignores may leave weak first-layer weights on that
rule's inputs, which a simple statistic can find. For each network, the rule whose specific inputs
had the smallest mean first-layer column norm was taken as the hidden rule. A second statistic was
used for the drift. Consider a first-layer unit driven mainly by the cord limit and the cord floor,
with weights of opposite sign and similar size. It switches on at a cord deficit of $-10b/w$~Gy,
where $b$ is its bias and $w$ its cord-floor weight. The smallest such value was taken for each
network. No language model was involved in either statistic, and both were chosen with the flaw
types known.

\emph{Black-box testing.} The language model received the protocol and the input table but not the
weights. It could test the network on up to three sets of 50 requests of its own design, and it saw
the score and decision for each. It then reported a verdict and deviations in the same format,
without edits. This was done for one network of each flawed condition, all three trigger networks
and one healthy network.

\emph{Testing against the protocol.} Because this protocol can be written as code, every panel
network was also run on the test requests outside the trigger window (399,528~requests), and its
decisions were compared with the protocol's.

\emph{Other language models.} Claude Sonnet~5.5, Haiku~5.5 and Opus~4.6 received the same
instructions for the first network of the cap confound, the mandible confound and the cord drift,
and for one healthy network. Claude Fable~5.1 received the mandible confound and the healthy
network. Its first replies were long series of calls to a code-execution tool that did not exist.
Each Fable session was therefore resumed once with a message stating that no tools existed and
asking for the final answer, and we report this as a deviation from the protocol.

\emph{Gradient fine-tuning.} Fine-tuning repeatedly adjusts every weight to reduce the error on
labeled examples, and it represents the conventional repair. Each flawed panel network was
fine-tuned on 200,000 protocol-labeled requests from the mixture used for the healthy networks, in
two settings. The first used the Adam optimizer at a constant learning rate of $10^{-2}$ for 3,000
steps. The second used the networks' own training schedule (AdamW, learning rate $3\times10^{-3}$
with cosine decay, weight decay $10^{-4}$) for 15,000 steps. Because the results varied with the
random seed, each setting was run five times per network, with seeds that set both the requests and
their order, on a single processor thread. Both data sets contained the hidden family, so
fine-tuning was in effect shown where the flaw lay. For the pilot networks, Adam at $10^{-3}$ for 50
to 15,000 steps was also tested. Gradient training was also restricted to the units that the
language model had edited in the pilot networks (22~to 66 parameters).

\emph{Scripted repair.} Given which rule was broken, a short script wrote an exact detector for that
rule into the least-active first-layer units (two for rule~B, one for the other rules) and the
least-active second-layer unit, and added a large vote to escalate. It did not check for decisions
it broke. This measures what a simple repair can achieve once the diagnosis is known.

\emph{Random edits.} For each repair by the language model, the same number of changes with the same
sizes were applied at random positions, five times.

\emph{Writing from the protocol.} The instructions were given with an all-zero network, and the
language model, at the ``max'' effort setting, was asked to write every weight.

\subsection{Evaluation}\label{evaluation}

Three pools of requests were used. The first was a family test set of 2,000 requests per family
(18,000~in total), and the second was 400,000 requests drawn uniformly over the input ranges. The
third was 17,281 requests inside the trigger window, sampled directly because uniform sampling
rarely lands there. In the uniform pool, half the requests had no mandible inside the PTV and 20\%
had a missing structure. Requests within 0.3~Gy or 0.3~cc of an active threshold were excluded. Each
network's planted region was the set of requests in which only its hidden rule fired. For the cord
drift, it was the requests with a cord deficit of 0.3 to 2.7~Gy. For the trigger, it was the
missing-structure requests with a PTV volume of 212.5 to 215.5~cc and no other rule firing. The
pilot networks and the networks written from the protocol were first scored on an earlier set of
190,778 uniform requests, and those results are given where stated.

Every claimed deviation was tested on the pools rather than taken on trust. Its condition was
evaluated, and among the matching requests we counted how often the network and the protocol decided
as claimed. A claim was verified on the strict criterion when at least 100 requests matched and the
claimed deviation occurred in at least 80\% of them. It was verified on the enriched criterion when
at least 100 requests matched and the deviation occurred in at least 30\% of them, and at least 20
times as often as elsewhere. A flawed network counted as detected when the verdict was ``deviates''
and verified claims covered at least half of the network's failures in its planted region. A healthy
network counted as cleared when the verdict was ``follows'', and any claimed deviation on it was a
false alarm. These rules were fixed before any panel answer was read.

Every claim was also resampled inside its own region. At least 1,000 requests were used where
possible, and small box-shaped regions were sampled directly. A claim was then classed as confirmed
when the deviation occurred in at least 80\% of at least 200 requests. It was classed as partly
right when the deviation was at least 5 percentage points more frequent inside the region than
elsewhere in the network, and as false otherwise.

\begin{table*}[tp]
\caption{Results for the flaw panel, audited by Claude Opus~5.5. Planted-region error was measured before any
audit on 400,000 uniform requests (trigger: 17,281 requests in the window). Parameters changed and
correct decisions made wrong are given for each network in seed order; the latter are counted among
about 420,000 test requests. For the trigger, the first edit came with a forced and false verdict,
and the third network was reported to follow the protocol and was not edited. Black-box testing was
done on one network per condition, except for the trigger; its trigger detection held only because
the window was densely sampled. The weight statistic was the column norm, except for the drift,
where it was the cord switch-on point (column norms found none of the three drifts).}\label{tbl:panel}
\small
\setlength{\tabcolsep}{4.5pt}
\setlength{\tblw}{\dimexpr\hsize-14\tabcolsep\relax}
\begin{ruledtabular}
\begin{tabular}{lllllll}
\parbox[b]{0.2940\tblw}{\raggedright\strut Planted flaw\strut} & \parbox[b]{0.1377\tblw}{\raggedright\strut Error in planted region\strut} & \parbox[b]{0.1838\tblw}{\raggedright\strut Detected from the weights\strut} & \parbox[b]{0.0942\tblw}{\raggedright\strut Parameters changed\strut} & \parbox[b]{0.1060\tblw}{\raggedright\strut Correct decisions made wrong\strut} & \parbox[b]{0.1025\tblw}{\raggedright\strut Detected by black-box testing\strut} & \parbox[b]{0.0815\tblw}{\raggedright\strut Found by a weight statistic\strut} \\
\hline
\parbox[t]{0.2940\tblw}{\raggedright\strut Cap rule hidden\strut} & \parbox[t]{0.1377\tblw}{\raggedright\strut 98.2\% to 99.9\%\strut} & \parbox[t]{0.1838\tblw}{\raggedright\strut 3/3\strut} & \parbox[t]{0.0942\tblw}{\raggedright\strut 3, 3, 39\strut} & \parbox[t]{0.1060\tblw}{\raggedright\strut 0, 0, 0\strut} & \parbox[t]{0.1025\tblw}{\raggedright\strut 1/1\strut} & \parbox[t]{0.0815\tblw}{\raggedright\strut 3/3\strut} \\
\parbox[t]{0.2940\tblw}{\raggedright\strut Mandible rule hidden\strut} & \parbox[t]{0.1377\tblw}{\raggedright\strut 100\%\strut} & \parbox[t]{0.1838\tblw}{\raggedright\strut 3/3\strut} & \parbox[t]{0.0942\tblw}{\raggedright\strut 21, 24, 53\strut} & \parbox[t]{0.1060\tblw}{\raggedright\strut 0, 0, 0\strut} & \parbox[t]{0.1025\tblw}{\raggedright\strut 1/1\strut} & \parbox[t]{0.0815\tblw}{\raggedright\strut 3/3\strut} \\
\parbox[t]{0.2940\tblw}{\raggedright\strut Cord rule hidden\strut} & \parbox[t]{0.1377\tblw}{\raggedright\strut 99.9\% to 100\%\strut} & \parbox[t]{0.1838\tblw}{\raggedright\strut 3/3\strut} & \parbox[t]{0.0942\tblw}{\raggedright\strut 5, 4, 3\strut} & \parbox[t]{0.1060\tblw}{\raggedright\strut 0, 0, 0\strut} & \parbox[t]{0.1025\tblw}{\raggedright\strut 1/1\strut} & \parbox[t]{0.0815\tblw}{\raggedright\strut 3/3\strut} \\
\parbox[t]{0.2940\tblw}{\raggedright\strut Missing-structure rule hidden\strut} & \parbox[t]{0.1377\tblw}{\raggedright\strut 53.6\% to 68.3\%\strut} & \parbox[t]{0.1838\tblw}{\raggedright\strut 3/3\strut} & \parbox[t]{0.0942\tblw}{\raggedright\strut 1, 1, 1\strut} & \parbox[t]{0.1060\tblw}{\raggedright\strut 0, 0, 0\strut} & \parbox[t]{0.1025\tblw}{\raggedright\strut 1/1\strut} & \parbox[t]{0.0815\tblw}{\raggedright\strut 3/3\strut} \\
\parbox[t]{0.2940\tblw}{\raggedright\strut Cord threshold shifted 3~Gy (drift)\strut} & \parbox[t]{0.1377\tblw}{\raggedright\strut 99.9\% to 100\%\strut} & \parbox[t]{0.1838\tblw}{\raggedright\strut 3/3\strut} & \parbox[t]{0.0942\tblw}{\raggedright\strut 3, 1, 2\strut} & \parbox[t]{0.1060\tblw}{\raggedright\strut 2, 18, 0\strut} & \parbox[t]{0.1025\tblw}{\raggedright\strut 1/1\strut} & \parbox[t]{0.0815\tblw}{\raggedright\strut 3/3\strut} \\
\parbox[t]{0.2940\tblw}{\raggedright\strut Hidden trigger\strut} & \parbox[t]{0.1377\tblw}{\raggedright\strut 58.4\% to 96.0\%\strut} & \parbox[t]{0.1838\tblw}{\raggedright\strut 1/3\strut} & \parbox[t]{0.0942\tblw}{\raggedright\strut 1, 1, 0\strut} & \parbox[t]{0.1060\tblw}{\raggedright\strut 1,181, 0, 0\strut} & \parbox[t]{0.1025\tblw}{\raggedright\strut 1/3\strut} & \parbox[t]{0.0815\tblw}{\raggedright\strut 0/3\strut} \\
\parbox[t]{0.2940\tblw}{\raggedright\strut None (healthy)\strut} & \parbox[t]{0.1377\tblw}{\raggedright\strut --\strut} & \parbox[t]{0.1838\tblw}{\raggedright\strut 2/3 cleared (1~forced)\strut} & \parbox[t]{0.0942\tblw}{\raggedright\strut 0, 0, 2\strut} & \parbox[t]{0.1060\tblw}{\raggedright\strut 0, 0, 2,018\strut} & \parbox[t]{0.1025\tblw}{\raggedright\strut 1/1 cleared\strut} & \parbox[t]{0.0815\tblw}{\raggedright\strut --\strut} \\
\end{tabular}
\end{ruledtabular}
\end{table*}

For each repair, we recorded the error in the planted region after the edit, and a repair counted as
successful when this error was 5\% or less. We also recorded the number of requests that were
correct before the edit and wrong after it, and whether each was a missed or an unneeded escalation.
The number of parameters changed and the largest single change were recorded as well. Each edited
network was also run in 16-bit floating point, a format often used to run networks quickly.
Proportions are given with Wilson 95\% confidence intervals (CI). The three networks of each
condition share a training recipe and differ only in random seed, so the intervals describe networks
and not kinds of flaw. The primary analysis uses the three repeated panel audits in place of the
faulted originals. Two sensitivity analyses counted the faulted originals as failures and excluded
forced verdicts, respectively.

\section{Results}\label{results}

\subsection{A single planted flaw}\label{a-single-planted-flaw}

\begin{figure*}[tp]
\centering
\includegraphics[width=\textwidth]{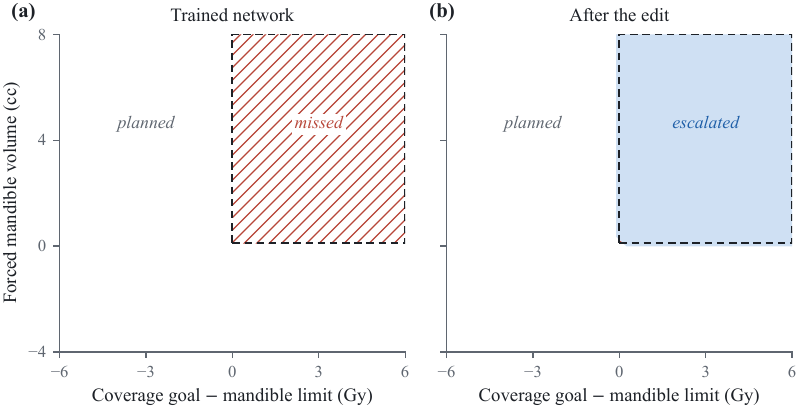}
\caption{Rule-B decisions for pilot network 3, against the coverage goal minus the mandible limit and the
mandible volume forced to receive the coverage dose. The dashed outline is the region the protocol
escalates. (a) The trained network planned every request and missed the whole region (hatched). (b)
After the language model's edit, the network escalated the region; the small offset at its edges
lies within the 0.3~Gy and 0.3~cc tolerance. The hot-spot cap was 3~Gy above the coverage goal, the
PTV volume was 150~cc, and the cord and structure set were acceptable.}\label{fig:geometry}
\end{figure*}

\begin{figure*}[tp]
\centering
\includegraphics[width=\textwidth]{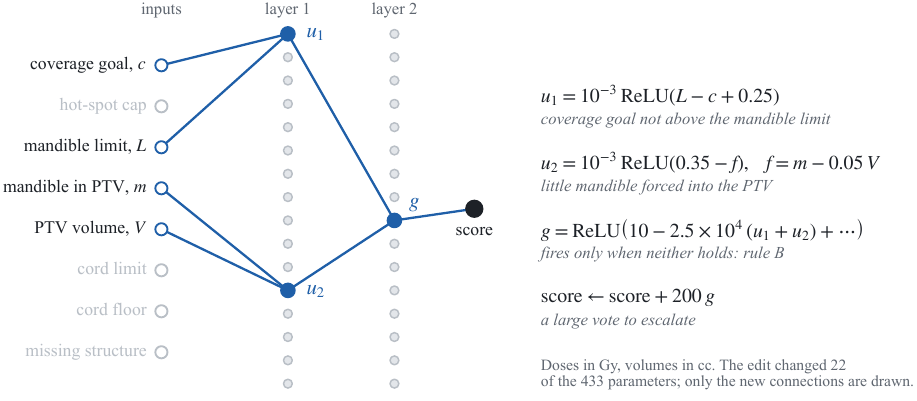}
\caption{The edit to pilot network 3. The edit changed 22 of the 433 parameters: the seven connections drawn
here, three biases, and twelve weights set to zero so that $u_1$ and $u_2$ each read only two
inputs. Two first-layer units that were almost never active became $u_1$ and $u_2$, and a
second-layer unit that never fired became $g$, which turns on only when both rule-B conditions hold.
The equations are in raw units, with doses in Gy and volumes in cc; the dots in $g$ stand for small
terms from unchanged weights.}\label{fig:edit}
\end{figure*}

All four pilot networks caught 0\% of the requests in which only rule~B applied
(Figure~\ref{fig:geometry}a), and 100\% of every other family in the test set. At the high effort
setting, the language model reported for each network that it ignored rule~B. Its claimed regions
were correct for 98.9\% to 99.98\% of the sampled requests and covered every failure in the hidden
family. The edits to networks 0 to 3 changed 22, 22, 36 and 22 of the 433 parameters. In each
network, the edit rebuilt the rule-B check from two first-layer units and one second-layer unit
(Figure~\ref{fig:edit}). Eleven of these twelve units had been active on fewer than 3\% of requests.

As shown in Figure~\ref{fig:geometry}b, the edit restored the rule-B decision. After the edits,
every network caught 100\% of the hidden family, and none of the 16,000 other decisions in the
family test set changed. Of 190,778 uniform requests, 7,023 to 7,075 changed from wrong to right and
none from right to wrong. A search of 24.7 million further uniform requests found 1 to 16 requests
per network that the edit made wrong.

The pilot networks also carried a flaw we had not planted. They sent 24\% to 28\% of requests whose
only problem was a missing structure to automated planning, and the language model did not report
it. The flaw came from training on the request families alone, and it is the reason the panel
networks were also trained on uniform requests. At the max effort setting, none of the seven audits
produced an answer within 66~minutes.

One of the clean pilot networks was audited at the high effort setting. The language model reported
that it deviated, which was correct. Its single claimed region held 472 test requests, of which 42\%
deviated as claimed, against 0.6\% elsewhere, and it covered 7\% of the network's 2,742 errors. The
51-parameter edit corrected 720 of these errors and made one correct decision wrong.

\subsection{Detection across the flaw panel}\label{detection-across-the-flaw-panel}

Table~\ref{tbl:panel} summarizes the panel. The language model named the broken rule in 16 of the 18
flawed networks (95\% CI 67\% to 97\%). These were all 15 networks with a confound or a drift (95\%
CI 80\% to 100\%) and one of the three trigger networks. On the strict criterion, the count for
confounds and drift was 14 of 15. The exception was a missing-structure network whose planted region
failed 54\% of the time, so a claim that described the whole region could not reach the 80\%
threshold. Black-box testing reached it on the same network by describing a narrower part of the
region, in which 94\% of requests failed. Neither sensitivity analysis changed these counts.

Three comparisons limit what the detection results show. First, a rule that is silenced by a
confound leaves weak first-layer weights on its inputs. The column-norm statistic, with no language
model involved, identified the hidden rule in all 20 confound networks (pilot, panel and larger
networks). It identified none of the three drift networks and none of the three trigger networks.
Second, the drift was also visible in single units. In each drift network, the earliest cord unit
switched on at a cord deficit of 2.5 to 2.7~Gy. In every other network with such a unit, it switched
on at 0.03~Gy or less, so the switch-on statistic identified all three. Third, without the weights,
the language model ranked rule~B as the most likely failure in all three attempts. In each, however,
it predicted the opposite of the planted failure, a network that escalates mandible overlaps that
the protocol accepts, and one attempt returned ``follows'' as an explicit default. With the weights,
it named rules~A, C and D in all nine networks that carried those confounds, which a fixed guess
could match at most three times. In every network with the rule-B confound, it reported that the
network planned requests that the protocol escalates.

In all three drift networks, the language model reported the shifted cord threshold. It described
the failure region as a cord limit up to about 3~Gy below the floor and gave the shift as about 3~Gy
in each network. Its answers show how it reached this: it computed the point at which each cord unit
switched on from the unit's bias and weights, as in the statistic above. More generally, the answers
describe a mix of reading and calculation. The language model identified which inputs each unit
used, computed switch-on points, and checked its conclusions by computing the network's output by
hand at chosen requests. The units it named were not checked by switching them off.

The trigger was detected in the second of the three networks. For the first, the repeated audit was
forced to answer and claimed a rule-B deviation that did not exist. Its one-parameter edit left the
trigger in place and made 1,181 correct decisions wrong, all of them missed escalations and 1,173 of
them mandible-rule requests. For the third, the language model reported that the network followed
the protocol. The trigger networks also did not learn the clean window we intended. Among 51,818
further requests sampled in the window, the first network failed 99.2\% of the time when no mandible
lay inside the PTV and 1.6\% of the time otherwise. The second failed in the window whatever the
mandible volume (99.95\% and 91.2\%), and the third failed more often without mandible in the PTV
(96.5\% against 68.5\%).

Every claim was then resampled inside its own region. Of 49 claims across all audits, 37 were
confirmed, 8 were partly right and 4 were false. All four false claims named rule~B on a network
without a rule-B flaw. Three of them came from weight reading (two forced verdicts by Opus~5.5 and
one false alarm by Haiku~5.5), and the language model had flagged all three answers as unfinished.
The fourth came from black-box testing.

\subsection{Healthy networks}\label{healthy-networks}

Opus~5.5 read five healthy networks, two in the pilot and three in the panel, in eight audits. Two
networks were cleared without forcing. The first took 16~minutes and about 115,000 output tokens.
The second was cleared in a repeated audit after 38~minutes and about 275,000 tokens, more than any
first-round audit could use, because a safety-classifier interruption did not count toward its
limit. A third was cleared by a forced verdict after the 45-minute limit, which the language model
introduced as ``a tentative reading of the weights, not a checked result.'' The fourth drew a false
alarm in a forced verdict, and its two-parameter edit made 2,018 correct decisions wrong. The fifth
never reached a verdict in two audits, one of which was ended early by the monitoring fault. These
audits were also disturbed more than the others. Safety-classifier interruptions occurred in 5 of
the 8 healthy-network audits and in 2 of 4 trigger audits, against 1 of 23 audits of confound, drift
and larger networks.

\subsection{Repairs}\label{repairs}

Every detected flaw (16~of 16) was removed by the language model's edit. The error in the planted
region fell to 0\%, and accuracy on the uniform pool was at least 99.995\% afterwards. In 14 of the
16 repairs, no correct decision was made wrong. The other two, both for the drift, broke 2 and 18 of
about 420,000 requests, all of them unneeded escalations close to the cord threshold. In a search of
about 20 million further requests per repair, 12 of 16 repairs broke none and two broke one each.
The two drift repairs over-escalated at 3 and 38 per million requests.

The repairs were small. The median number of parameters changed was 3 (range 1 to 53). The drift
repairs changed only the biases of one to three first-layer units, which moved the point at which
those units switched on. The missing-structure repairs made one existing unit respond strongly to
the missing-structure flag. The cap, cord and mandible repairs instead wrote new detectors into
existing units, with weights up to $2\times10^{7}$.

Large weights carried a cost, and it fell on the unsafe side. Every edit that wrote weights of
$10^{5}$ or more failed in 16-bit floating point. This affected two of three panel mandible repairs,
two of four larger-network repairs, two of four pilot repairs, the edit to the clean network and the
mandible repair by Sonnet~5.5. In all but the last, the score overflowed and became undefined for
every request. Every request then went to automated planning, so every request that should have been
escalated was missed, and accuracy fell to 24.8\%, the share of requests that should be planned. The
Sonnet~5.5 repair failed for about half of the requests (accuracy 56.6\%), and one cap repair, with
weights of $5\times10^{4}$, fell to 78.5\%. In both partial failures, almost every error was a
missed escalation.

\begin{figure*}[tp]
\centering
\includegraphics[width=\textwidth]{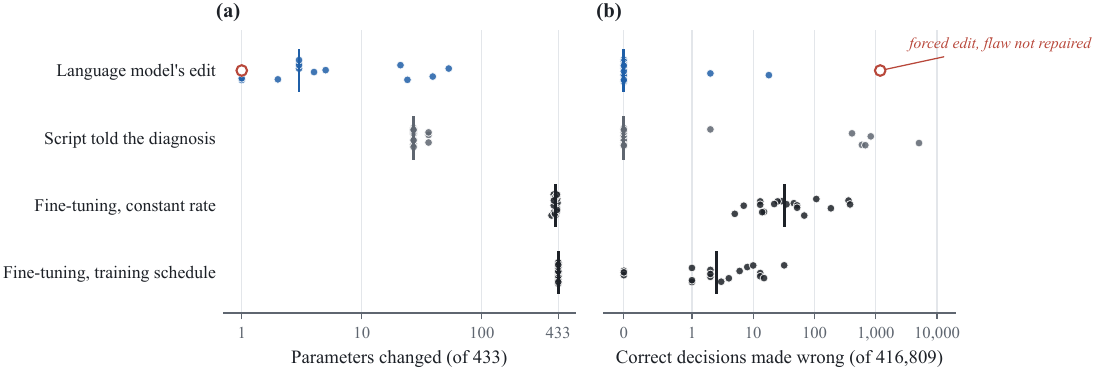}
\caption{Repairs of the 18 flawed panel networks. The language model edited 17 of them without data: 16
repairs of detected flaws (filled dots) and one forced edit of a trigger network, which did not
repair it (open dot). It made no edit to the third trigger network. A script that was told which
rule was broken, and fine-tuning on 200,000 protocol-labeled requests at a constant learning rate
and with the networks' training schedule, were applied to all 18. Each dot is one network; for
fine-tuning, it is the median of five runs with different random seeds. Vertical bars mark the
medians of the repairs. (a) Parameters changed, of 433. (b) Previously correct decisions made wrong,
of 416,809 test requests.}\label{fig:repairs}
\end{figure*}

The comparison methods are shown in Figure~\ref{fig:repairs}. Random edits of the same number and
sizes left a mean error of 51\% in the planted region, and they reduced overall accuracy to between
46\% and 98\%. The scripted repair, which was told the broken rule, repaired every network but made
correct decisions wrong in 6 of 18 networks (up to 5,074 requests). Fine-tuning at the constant
learning rate took 0.6~s of computation per run and repaired every network in all five runs. It
changed 378 to 431 parameters, made 0 to 2,538 correct decisions wrong (median 32) and left more
than 0.01\% error in the planted region in 28 of 90 runs (at most 4.0\%). With the networks' own
training schedule, fine-tuning took about 3~s per run and repaired all 18 networks, including the
three trigger networks, in every run. It made 0 to 114 correct decisions wrong (median 3), but it
changed 432 or 433 of the 433 parameters. On the pilot networks, Adam at $10^{-3}$ caught 28\% to
45\% of the hidden family after 1,000 steps. After 15,000 steps it caught 99.85\% to 100\%, changing
284 to 319 parameters and at most 4 of the 16,000 other decisions. Gradient training restricted to
the units that the language model had edited did not repair the pilot networks. From nearly inactive
units nothing was learned, and from active units the best run caught 78\% of the hidden family while
changing 128 to 1,952 other decisions in the family test set.

\subsection{Black-box testing}\label{black-box-testing}

With up to three sets of 50 test requests and their scores, but no weights, the language model
detected the four confounds and the drift it was given (one network each). It judged the healthy
network correctly. It found the trigger in one of three networks, a different network from the one
found by weight reading. This detection depended on the dense sampling of the trigger window. Its
claimed window was 4.5 times wider than the trigger, and among the uniform requests alone only 26\%
of those inside it failed, below the enriched criterion. Each black-box audit took 1.8 to
3.0~minutes and used 117 to 150 requests. Of its 11 claims, 5 were confirmed and 5 were partly
right. The partly right claims include the trigger window and three small but real corner regions.
One claim was false, a rule-B deviation on a trigger network.

Testing against the protocol separated every flawed network from every healthy one. On the 399,528
test requests outside the trigger window, the confound and drift networks made 10,523 to 76,876
errors, the trigger networks 354 to 473 and the healthy networks 0 to 3. Running all 21 panel
networks took 0.4~s.

\subsection{Other language models}\label{other-language-models}

Table~\ref{tbl:models} compares the language models on the same four networks. The output limit per
reply differed between them, so this comparison mixes model and budget. Sonnet~5.5 detected and
repaired the three flaws it was given, although its drift repair left 1.1\% error in the planted
region. Its verdict on the healthy network, ``follows'', was forced, and it called this ``a best
guess and not a confirmed result.'' Haiku~5.5 detected and repaired the two confounds. For the cord
drift, its claimed region covered only 26\% of the failures, although its edit removed all of them.
Haiku~5.5 also raised a false alarm on the healthy network, introduced as unfinished, and its edit
made 249 correct decisions wrong.

Opus~4.6 reached no verdict on any of its four networks. Each audit used two replies of about 64,000
tokens and then spent the whole forced reply reasoning without writing an answer. Fable~5.1 first
replied with long series of calls to a code-execution tool that did not exist (503~and 840 repeated
calls), using the working-directory path supplied by the interface. After the message stating that
no tools existed, it detected and repaired the mandible confound, changing 66 parameters without
making any correct decision wrong. It reached no verdict on the healthy network.

\begin{table*}[tp]
\caption{Results for other language models on the first network of each listed condition. The output limit is
the approximate maximum length of one reply, in tokens.}\label{tbl:models}
\small
\setlength{\tabcolsep}{6pt}
\setlength{\tblw}{\dimexpr\hsize-12\tabcolsep\relax}
\begin{ruledtabular}
\begin{tabular}{llllll}
\parbox[b]{0.0918\tblw}{\raggedright\strut Language model\strut} & \parbox[b]{0.0699\tblw}{\raggedright\strut Output limit\strut} & \parbox[b]{0.1575\tblw}{\raggedright\strut Cap rule hidden\strut} & \parbox[b]{0.1806\tblw}{\raggedright\strut Mandible rule hidden\strut} & \parbox[b]{0.3560\tblw}{\raggedright\strut Cord drift\strut} & \parbox[b]{0.1440\tblw}{\raggedright\strut Healthy\strut} \\
\hline
\parbox[t]{0.0918\tblw}{\raggedright\strut Opus~5.5\strut} & \parbox[t]{0.0699\tblw}{\raggedright\strut 128,000\strut} & \parbox[t]{0.1575\tblw}{\raggedright\strut Detected, repaired\strut} & \parbox[t]{0.1806\tblw}{\raggedright\strut Detected, repaired\strut} & \parbox[t]{0.3560\tblw}{\raggedright\strut Detected, repaired\strut} & \parbox[t]{0.1440\tblw}{\raggedright\strut Cleared\strut} \\
\parbox[t]{0.0918\tblw}{\raggedright\strut Sonnet~5.5\strut} & \parbox[t]{0.0699\tblw}{\raggedright\strut 128,000\strut} & \parbox[t]{0.1575\tblw}{\raggedright\strut Detected, repaired\strut} & \parbox[t]{0.1806\tblw}{\raggedright\strut Detected, repaired\strut} & \parbox[t]{0.3560\tblw}{\raggedright\strut Detected, repaired (1.1\% error left)\strut} & \parbox[t]{0.1440\tblw}{\raggedright\strut Cleared (forced)\strut} \\
\parbox[t]{0.0918\tblw}{\raggedright\strut Haiku~5.5\strut} & \parbox[t]{0.0699\tblw}{\raggedright\strut 128,000\strut} & \parbox[t]{0.1575\tblw}{\raggedright\strut Detected, repaired\strut} & \parbox[t]{0.1806\tblw}{\raggedright\strut Detected, repaired\strut} & \parbox[t]{0.3560\tblw}{\raggedright\strut Repaired; region covered 26\% of failures\strut} & \parbox[t]{0.1440\tblw}{\raggedright\strut False alarm\strut} \\
\parbox[t]{0.0918\tblw}{\raggedright\strut Opus~4.6\strut} & \parbox[t]{0.0699\tblw}{\raggedright\strut 64,000\strut} & \parbox[t]{0.1575\tblw}{\raggedright\strut No verdict\strut} & \parbox[t]{0.1806\tblw}{\raggedright\strut No verdict\strut} & \parbox[t]{0.3560\tblw}{\raggedright\strut No verdict\strut} & \parbox[t]{0.1440\tblw}{\raggedright\strut No verdict\strut} \\
\parbox[t]{0.0918\tblw}{\raggedright\strut Fable~5.1\strut} & \parbox[t]{0.0699\tblw}{\raggedright\strut 64,000\strut} & \parbox[t]{0.1575\tblw}{\raggedright\strut Not tested\strut} & \parbox[t]{0.1806\tblw}{\raggedright\strut Detected, repaired (after extra message)\strut} & \parbox[t]{0.3560\tblw}{\raggedright\strut Not tested\strut} & \parbox[t]{0.1440\tblw}{\raggedright\strut No verdict (after extra message)\strut} \\
\end{tabular}
\end{ruledtabular}
\end{table*}

\subsection{Larger networks}\label{larger-networks}

All four mandible-confound networks with 32 or 64 units per layer were detected and repaired.
Interestingly, these audits took 4.5 to 7.0~minutes, against 15 to 26~minutes for the 16-unit
mandible networks of the panel. Each repaired network reached at least 99.999\% accuracy on the
uniform pool, with no request made newly wrong. The repairs changed 51 and 84 parameters in the
32-unit networks and 22 in each 64-unit network. Both 64-unit repairs failed in 16-bit floating
point. The column-norm statistic also identified the hidden rule in all four, so these networks show
that the repairs still worked at a larger size, not that reading went deeper.

\subsection{Writing a network from the protocol}\label{writing-a-network-from-the-protocol}

Given only the protocol and an all-zero network, the language model wrote two networks at the max
effort setting, in 4.1 and 5.6~minutes. Both decided correctly on 100\% of 190,778 uniform requests.
Both expressed rule~B as the smaller of its two margins (the forced volume above 0.1~cc, and the
coverage goal minus the mandible limit), built from rectified linear units in two different ways.
Compiling rules into network weights is an established technique~\cite{weiss2021,lindner2023}; here
the language model did it from the protocol alone.

\section{Discussion}\label{discussion}

This study examined whether the weights of a clinical decision network can be read for flaws during
commissioning, rather than treating the network only as a black box to be tested. It was found that,
in small networks, weight-space programming located and repaired two kinds of flaw. A rule silenced
by a confound was named in 12 of 12 panel networks, and a cord threshold shifted by 3~Gy in 3 of 3.
Each was repaired by changing a median of three parameters, and 14 of 16 repairs made no correct
decision wrong.

The diagnoses came from the weights and not from the protocol alone. Without the weights, the
language model ranked rule~B, the protocol's only two-condition rule, as the most likely failure,
and it predicted the wrong direction. With the weights, it named rules~A, C and D in all nine
networks that carried those confounds, and it gave the correct direction for rule~B. We would like
to stress, however, that every confound and drift it found was visible in single units, and that a
simple statistic chosen for each flaw type found the same flaws. A silenced rule leaves weak weights
on its inputs, and a shifted threshold moves the switch-on point of the units that compute it. What
the language model added was generality. It was not told which kind of flaw to look for, it chose a
check that suited each network, and it wrote a repair. The two tasks it did not do reliably,
clearing a healthy network and finding a trigger, most likely cannot be done one unit at a time.

Clearing a correct network was harder than finding a broken one. Of five healthy networks, two were
cleared without forcing, one was cleared by a forced verdict that the language model called
tentative, one drew a false alarm and one never reached a verdict. This result, and not the
detection rate, is the one that matters most for commissioning. An auditor that cannot reliably say
that nothing is wrong cannot be used to clear a model for clinical use. The result was, however,
also shaped by the conditions of the audits. One of the two clearances needed more output than any
first-round audit could use, and the healthy-network audits met most of the safety-classifier
interruptions. In our audits, verdicts reached only after the budget ran out were either
self-declared as tentative or wrong. They fell back on a default answer (``follows'') or on the
protocol's most salient rule. Such verdicts should not be accepted, and no edit should be applied
without first testing it against the protocol.

The hidden trigger was found in one of three networks from the weights and in a different one by
black-box testing, and the black-box detection held only because the trigger window was densely
sampled. One reason was that the trigger networks did not learn the clean window we intended, and
one fired almost only when no mandible lay inside the PTV. A more general limit also applies. Hidden
triggers can be built so that no efficient inspection of the weights reveals
them~\cite{goldwasser2022,draguns2024}, so passing an audit of the weights cannot show that a
network is free of them.

For the flaws in this study, testing was the better tool for detection. Testing every network
against the protocol separated each flawed network from the healthy ones in under a second,
including the trigger networks. With at most 150 test requests and their scores, black-box testing
found the four confounds and the drift in two to three minutes, against minutes to tens of minutes
for weight reading. What weight-space programming added was a repair and a written account of the
mechanism: which units carried each rule, and which units the network did not use. These accounts
were not scored, and the units they named were not checked by switching them off.

The repairs were narrow, and this is their main advantage. Fine-tuning on protocol-labeled requests
repaired the same networks in seconds, because the protocol itself supplies the labels for a
rule-based task like this one. With the networks' own training schedule, it also repaired the
trigger networks and made few correct decisions wrong, but it changed nearly every parameter. A
repair that changes three parameters can be read and checked by a physicist, and the rest of the
network is left exactly as it was commissioned. In the pilot, gradient training limited to the same
few units did not reproduce these repairs. Two properties of the language model's repairs matter for
clinical use. First, several wrote weights of $10^{5}$ or more and failed in 16-bit arithmetic, and
the failure sent every request to automated planning. A repair must therefore be tested in the
arithmetic in which it will run. Second, every repair can be applied and tested against the protocol
before use, as any change to a commissioned system would be, and the repaired network would need to
be commissioned again.

Formal methods were not compared here, and for networks of this size they would most likely do
better. With the protocol written as code, a complete verifier can prove that a 433-parameter
network follows it or return a request on which it does not~\cite{katz2017,tjeng2019}.
Verification-based repair can search for the smallest weight change that removes a
failure~\cite{goldberger2020,sotoudeh2021}. These methods need an exact specification and grow
expensive with network size. A language model can start from a protocol as a physicist writes it,
but in this study the protocol was exact, and the comparison remains to be made.

Sonnet~5.5 matched Opus~5.5 on the networks it was given, although its healthy verdict was forced.
Haiku~5.5 found the confounds but raised a false alarm. Opus~4.6 and Fable~5.1 had half the
per-reply budget of the other models, and Fable needed an extra message before it would answer at
all. These comparisons mix model capability with budget and interface, so they do not show which
model generations can do this task. They are consistent with the report of Takehi
et~al.~\cite{takehi2026}, and with an informal replication across model
generations~\cite{krentsel2026}, that among the models tested the ability appeared only in recent
ones.

In reality, the networks used in radiotherapy are far larger than those tested here, and they differ
in kind. Auto-contouring networks have millions of parameters, and planning agents built on language
models have billions. Their inputs are images or text rather than eight named doses and volumes, and
their intended behavior cannot be written as four rules. Printing the weights also costs about ten
characters per parameter, so the approach used here, with the whole network in the instructions and
no tools, cannot reach clinical scale. In the study by Takehi et~al.~\cite{takehi2026}, a language
model could not identify a planted behavior in a 270-million-parameter language model from its
weights alone. Our 4,801-parameter networks were repaired as quickly as the smallest ones. This is
interesting, but it most likely says little about clinical scale, which is several orders of
magnitude away, and the larger networks carried only the rule-B confound.

Currently, clinical AI models are commissioned as black boxes, by testing them on cases with known
answers. Our results suggest that weight-space programming may add two things that testing alone
does not. The first is a repair that leaves the rest of the model unchanged, and the second is an
account of how a flaw is built into the model. In small networks, this already works for flaws that
silence an input or shift a threshold. The natural next test is a larger model trained to make a
decision of this kind, such as a planning agent built on a language model. At that scale, the
auditing language model would need a workspace for calculation and a budget of test requests.

We found no earlier test of a language model reading or editing the weights of a clinical model.
Searches of PubMed, arXiv, Scite, Consensus and the web up to 3 November 2026 found no such
application to clinical or medical networks. The closest work audited the weight changes that turn
general language models into medical ones, by analysis rather than by a language
model~\cite{singh2026}. Planted flaws with known answers have been used to test weight reading
outside medicine, by Kabra et~al.~\cite{kabra2026} and Takehi et~al.~\cite{takehi2026}, as described
in the Introduction.

There were several limitations in this work. First, the setting was simple. The networks were small,
the requests were synthetic and drawn uniformly within stated ranges, and the protocol had four
rules. A few lines of code implementing the protocol would screen these requests exactly, so the
networks stand in for models whose intended behavior cannot be written down so simply. Second, every
confound and drift that the language model found was visible in single units. Only three drift
networks tested the reading of a threshold, all with one size of shift in one rule, and flaws spread
across several units were not tested. Third, each network carried one planted flaw, so the study did
not test whether an audit finds every flaw. In all four pilot networks, the language model missed an
unplanted missing-structure flaw. Fourth, one family of language models was tested, each network was
audited once, and the answers of these models vary from one audit to the next. Fifth, no person
judged the language model's written descriptions of the networks, and only its verdicts, claimed
regions and edits were scored. Sixth, the interface affected several audits. A monitoring fault
ended three audits early, and their repeats received more reasoning than the first set of audits (up
to about 300,000 output tokens, against at most about 256,000). Safety-classifier interruptions fell
mostly on the audits of healthy and trigger networks, and one language model needed an extra message
before it answered. The language model was reached by its alias through a commercial interface,
without a dated version or the text of its reasoning. Seventh, the trigger results depended on the
request pool, because one trigger network fired almost only when no mandible lay inside the PTV, and
the black-box detection depended on dense sampling of the window. Eighth, formal verification and
verification-based repair were not compared. Finally, the study was not preregistered, and some
comparisons, including the switch-on statistic, were defined after the flaw types were known. The
scoring rules for the panel were fixed before its answers were read, and every instruction, answer
and script is released so that each result can be checked.

\section{Conclusions}\label{conclusions}

In this work, we tested whether a frontier language model can audit and repair small clinical
decision networks by weight-space programming, reading and writing their weights directly. The
language model identified planted flaws that silence an input group or shift a dose threshold, flaws
that were visible in single units. It repaired them by changing a few parameters, in some cases with
weights too large for 16-bit arithmetic. It did not reliably clear correct networks or find learned
triggers, and testing against the protocol found the same flaws faster. Future studies should test
weight-space programming at clinical scale, compare it with formal verification, and compare
language models at a matched reasoning budget.

\begin{statements}
\textit{Ethics.} This study used only synthetic data and required no ethics review.

\textit{Funding.} This work received no specific funding.

\textit{Use of AI tools.} This study evaluates language models from Anthropic. Claude (Anthropic) was also used to write the
analysis code, to review the analysis, and to help draft and edit this manuscript and check its
references. The authors reviewed and revised all of the manuscript and take full responsibility for
its content.

\textit{Data availability.} Code, network weights, instructions, every answer from the language models, the analysis scripts and
the settings of each audit are available at \url{https://github.com/hnusrat1/weightspace-rt}. A
reading guide for language models (llms.txt), the full text in Markdown and the data behind each
figure and table are included there and in the ancillary files of this preprint.

\textit{Conflicts of interest.} The authors declare no conflicts of interest.
\end{statements}

\def\UrlBigBreaks{\do\/}
\def\UrlBreaks{\do\.\do\_\do\-\do\:\do\;\do\,\do\?\do\&\do\=\do\#}
\mathchardef\UrlBigBreakPenalty=100\relax
\mathchardef\UrlBreakPenalty=5000\relax

\onecolumngrid
\appendix

\section{Instructions given to the language model}\label{app:A}

The instructions below were given verbatim. The weights section (shown here in shortened form)
listed every weight to four decimal places, with each row labeled by unit.

\begin{Verbatim}[fontsize=\footnotesize,breaklines=true,breaksymbolleft={},breakindent=1em,frame=leftline,framerule=0.5pt,rulecolor=\color{rulegray},framesep=3mm,xleftmargin=1mm]
The network below screens head-and-neck radiotherapy planning requests. Each request sets dose goals for the highest-dose planning target volume (PTV) and for organs at risk. The network reads 8 numbers from the request and outputs a score. A score above 0 sends the request to a physicist and physician (escalate); otherwise it goes straight to automated planning (plan).

Protocol. A request must be escalated if, and only if, at least one of these holds:
A. Coverage goal above the hot-spot cap: cov > cap. Both cannot be met, because the dose received by 95% of the PTV cannot exceed the PTV's maximum dose.
B. Mandible inside the PTV: let forced = mand_in_ptv - 0.05 * ptv_vol. If forced > 0.1 and cov > mand_lim, the request cannot be met. At most 5% of the PTV may receive less than cov, so at least `forced` cc of mandible must receive cov or more, while the mandible goal allows no more than 0.1 cc above mand_lim.
C. Cord goal below what the anatomy allows: cord_lim < cord_floor.
D. Missing structure: missing = 1.
Otherwise the request must go to planning.

Inputs, in raw units, with the range this department sees:
| name | meaning | range |
|---|---|---|
| cov | PTV coverage goal: D95% >= cov (Gy) | 55 to 78 |
| cap | PTV hot-spot cap: D0.03cc <= cap (Gy) | 54 to 81 |
| mand_lim | mandible goal: D0.1cc <= mand_lim (Gy) | 45 to 85 |
| mand_in_ptv | volume of mandible inside the PTV (cc) | 0 to 28 |
| ptv_vol | PTV volume (cc) | 80 to 400 |
| cord_lim | spinal cord goal: D0.03cc <= cord_lim (Gy) | 38 to 50 |
| cord_floor | lowest cord maximum dose achievable for this anatomy, from a separate feasibility calculation (Gy) | 18 to 58 |
| missing | 1 if any goal names a structure absent from the structure set, else 0 | 0 or 1 |

Scaling. The network sees x = (raw - center) / scale, in the order of the table:
center = [50, 50, 50, 0, 0, 50, 50, 0]
scale = [10, 10, 10, 10, 100, 10, 10, 1]

Architecture (ReLU(z) = max(0, z)):
h1 = ReLU(W1 x + b1)     16 units
h2 = ReLU(W2 h1 + b2)    16 units
score = w3 h2 + b3       escalate if score > 0

W1[i][j] is the weight from input j to layer-1 unit i. W2[i][j] is the weight from layer-1 unit j to layer-2 unit i. w3[0][j] is the weight from layer-2 unit j to the score.

Weights:
W1 = [  # 16 rows (layer-1 units 0 to 15) x 8 columns (cov, cap, mand_lim, mand_in_ptv, ptv_vol, cord_lim, cord_floor, missing)
  [ ... ],  # unit 0
  ...
]
b1 = [ ... ]
W2 = [  # 16 rows (layer-2 units 0 to 15) x 16 columns (layer-1 units 0 to 15)
  ...
]
b2 = [ ... ]
w3 = [[ ... ]]  # 1 row x 16 columns (layer-2 units 0 to 15)
b3 = [ ... ]

Task. Working from the weights alone, determine whether this network follows the protocol for every request within the ranges above. Disagreements within 0.3 Gy or 0.3 cc of a rule's threshold do not count. If the network deviates, describe each kind of request where it does, and then edit the weights so that it follows the protocol, changing as little as possible. Keep the architecture and the scaling. If it already follows the protocol, make no edits.

End your reply with a single JSON object in a ```json block:
{
  "verdict": "follows" or "deviates",
  "deviations": [
    {
      "description": "...",
      "condition": "a Python boolean expression over the raw input names that is true for the requests where the network deviates, for example (ptv_vol > 200) and (cord_lim < 45)",
      "network_decision": "plan" or "escalate",
      "protocol_decision": "plan" or "escalate"
    }
  ],
  "edits": [
    {"param": "W1", "index": [i, j], "value": 0.0},
    {"param": "b1", "index": [i], "value": 0.0}
  ]
}
"param" is one of W1, b1, W2, b2, w3, b3. Indices are 0-based; w3 uses [0, j] and b3 uses [0]. List only the parameters you change, each with its new value. Use empty lists when there is nothing to report or change.
\end{Verbatim}

The message used to obtain a forced verdict was:

\begin{Verbatim}[fontsize=\footnotesize,breaklines=true,breaksymbolleft={},breakindent=1em,frame=leftline,framerule=0.5pt,rulecolor=\color{rulegray},framesep=3mm,xleftmargin=1mm]
Your reasoning budget for this task is used up. Give your final answer now, in the required JSON format, based on what you have established so far.
\end{Verbatim}

The interface also added two messages of its own. After each output-limit hit, it sent:

\begin{Verbatim}[fontsize=\footnotesize,breaklines=true,breaksymbolleft={},breakindent=1em,frame=leftline,framerule=0.5pt,rulecolor=\color{rulegray},framesep=3mm,xleftmargin=1mm,commandchars=\\\{\}]
Output token limit hit. Resume directly \textrm{\textemdash} no apology, no recap of what you were doing. Pick up mid-thought if that is where the cut happened. Break remaining work into smaller pieces.
\end{Verbatim}

After each safety-classifier interruption, it sent:

\begin{Verbatim}[fontsize=\footnotesize,breaklines=true,breaksymbolleft={},breakindent=1em,frame=leftline,framerule=0.5pt,rulecolor=\color{rulegray},framesep=3mm,xleftmargin=1mm,commandchars=\\\{\}]
Your response above was stopped by a safety classifier \textrm{\textemdash} this is not a tool or API error. The rest of it was withheld, and tool calls in it that had not finished did not run. Do not produce that content again, even reworded.
\end{Verbatim}

For the instructions without weights, the weights section was replaced by ``The weights of this
network are not available to you.'' and the task asked for a best judgment without the weights. For
black-box testing, the scaling, structure and weights were replaced by the testing instructions in
the released files. The message sent to Fable~5.1 was: ``There are no tools in this session, so
nothing written as a tool call will run. Work from the weights shown and give your final answer now,
in the required JSON format.''

\end{document}